\documentclass[11pt]{article}
\usepackage{times}
\usepackage{geometry}
\usepackage[utf8]{inputenc}
\usepackage{enumitem,amssymb}
\usepackage{graphicx}
\usepackage{fancyhdr}
\usepackage{aas_macros}
\usepackage{mdframed} 
\usepackage{breakurl}
\usepackage[breaklinks]{hyperref}
\usepackage[superscript,biblabel]{cite}

\mdfdefinestyle{theoremstyle}{
innertopmargin=\topskip,}
\mdtheorem[style=theoremstyle]{lrptextbox}{}

\begin{document}

\begin{center}{\LARGE The Role of NewSpace in Furthering Canadian Astronomy}
\vspace{1em}

Aaron C.~Boley$^{1,*}$, 
David Kendall$^2$, 
Michael Byers$^1$, 
Frederic J.~Grandmont$^3$, 
Cameron Byers$^4$,
Jennifer Busler$^5$,
William MacDonald Evans$^6$,
Brett Gladman$^1$,
Tanya Harrison$^7$,
Catherine Johnson$^1$
\vspace{0.5em}

1) The University of British Columbia, 2) Former Chair of the UN Committee on the Peaceful Uses of Outer Space, 3) ABB Inc., 4) Gulf Islands Secondary School, 5) MDA, 6) Former President of the CSA, 7) Planet Labs
\vspace{0.5em}

$^*$aaron.boley@ubc.ca

\end{center}

\begin{center} Executive Summary \end{center}

Space has become more accessible, led by commercial actors taking an increasingly important role. SpaceX alone completed 21 launches in 2018, placing communication and navigation satellites into orbit, resupplying the ISS, and launching TESS. The Beresheet mission, while ultimately crashing on the Moon, was a major milestone because it was driven by a private company incorporated in Israel.  Such landings, whether successful or not, had previously only been attempted by major space-faring nations (Soviet Union/Russia, United States, EU, Japan, India, and China). A private American company (Moon Express) is planning a mission to the Moon with the International Lunar Observatory (ILO) as a payload.  If successful, it will be the first international facility in space organized by a non-profit global enterprise (ILOA).  A Canadian company is one of the primary contractors for this mission and the Canadian Space Agency (CSA) has signed a Memorandum of Understanding (MOU) with Moon Express to explore options for collaboration with the CSA and Canada's private space sector on technologies and payloads for missions to the Moon. In Low-Earth Orbit, new models for scientific study are also emerging.  When CSA funding for the satellite MOST ceased in 2014, the project was taken over by Microsatellite Systems Canada Inc., which continued operations using a pay-for-use model.  The UK mission Twinkle (2022 launch) is being developed and run by a private company that will offer photometric and spectroscopic observing capabilities under a similar pay-for-use model, with an anticipated strict proprietary data policy. Canada has formally joined the United States in the planned Lunar Orbital Platform-Gateway, and is planning to invest \$2 billion over the next 24 years.  If successful, the mission will see a space station and launch platform in orbit about the Moon that will facilitate human and robotic missions on the lunar surface.  Gateway will be open to both government and commercial actors.  It will also not be the only means for accessing the Moon and deep space.  Canadian-led deep space missions are possible, and we give an example of such a program, called Beacon.   As much as NewSpace presents opportunities, there are significant challenges that must be overcome, requiring engagement with policy makers to influence domestic and international space governance. Failure to do so could result in a range of long-lasting negative outcomes for science and space stewardship.  How will the Canadian astronomical community engage with NewSpace? What are the implications for NewSpace on the astro-environment, including Earth orbits, lunar and cis-lunar orbits, and surfaces of celestial bodies?   This white paper analyzes the rapid changes in space use and what those changes could mean for Canadian astronomers. Our recommendations are as follows: Greater cooperation between the astronomical and the Space Situational Awareness communities is needed. Build closer ties between the astronomical community and Global Affairs Canada (GAC).  Establish a committee for evaluating the astro-environmental impacts of human space use, including on and around the Moon and other bodies.  CASCA and the Tri-Council should coordinate to identify programs that would enable Canadian astronomers to participate in pay-for-use services at appropriate funding levels.  CASCA should continue to foster a relationship with CSA, but also build close ties to the private space industry.  Canadian-led deep space missions are within Canada's capabilities, and should be pursued.

\newpage

%


\begin{center}{\sc Scope of the White Paper}\end{center}

Old Space is a model in which space use is managed directly by states and their agencies, with large industrial companies serving as government contractors.  In contrast, NewSpace refers to the ever expanding role that private industry plays in space use, with programs, initiatives, and services led by commercial operators.  Indeed, the increasing number of non-state actors and the lowering cost of space access has permanently changed our daily lives. We are becoming ever more dependent on space services, such as navigation (sea, land, and personal), search and rescue, communications, weather forecasts, climate science, and environmental monitoring (e.g., fisheries and wildlife).  The broader utilization of space for education, entertainment, and art is also being developed.  NewSpace is often used as a reference to this new reality as well, which is how it is intended here. 

This white paper (WP) explores several prominent challenges of space use in the NewSpace era and highlights how those issues could impact Canadian astronomy.  The WP further discusses opportunities that NewSpace presents to the Canadian community, including leading lunar and deep space science programs. 

This work is complementary to that of Hutchings (2019)\cite{hutchings2019}, which provides an overview of Canadian space history and ongoing CSA projects, as well as the WP by Metchev et al.~(2019)\cite{metchev2019}, which discusses the use of smallsat technology for astronomy.  Furthermore, while there are a number of key challenges facing NewSpace, which involve technical, political, and legal hurdles, this WP will only reference those that will need immediate consideration for Canadian astronomy and space exploration.

The WP is organized as follows: A series of subtopics are presented as short summaries.  An overall discussion followed by key recommendations is presented at the end of the WP. When necessary to provide context, some discussion is also given in the subtopic summaries. 

\begin{center}{\sc Accessibility of Space}\end{center}

{\it The number of launch opportunities, the costs of those launches, and payload sharing are making space accessible to a wider group of people. }
\vspace*{0.5cm}

\begin{figure}[h]
    \centering
    \begin{minipage}{0.45\textwidth}
       \includegraphics[width=3in]{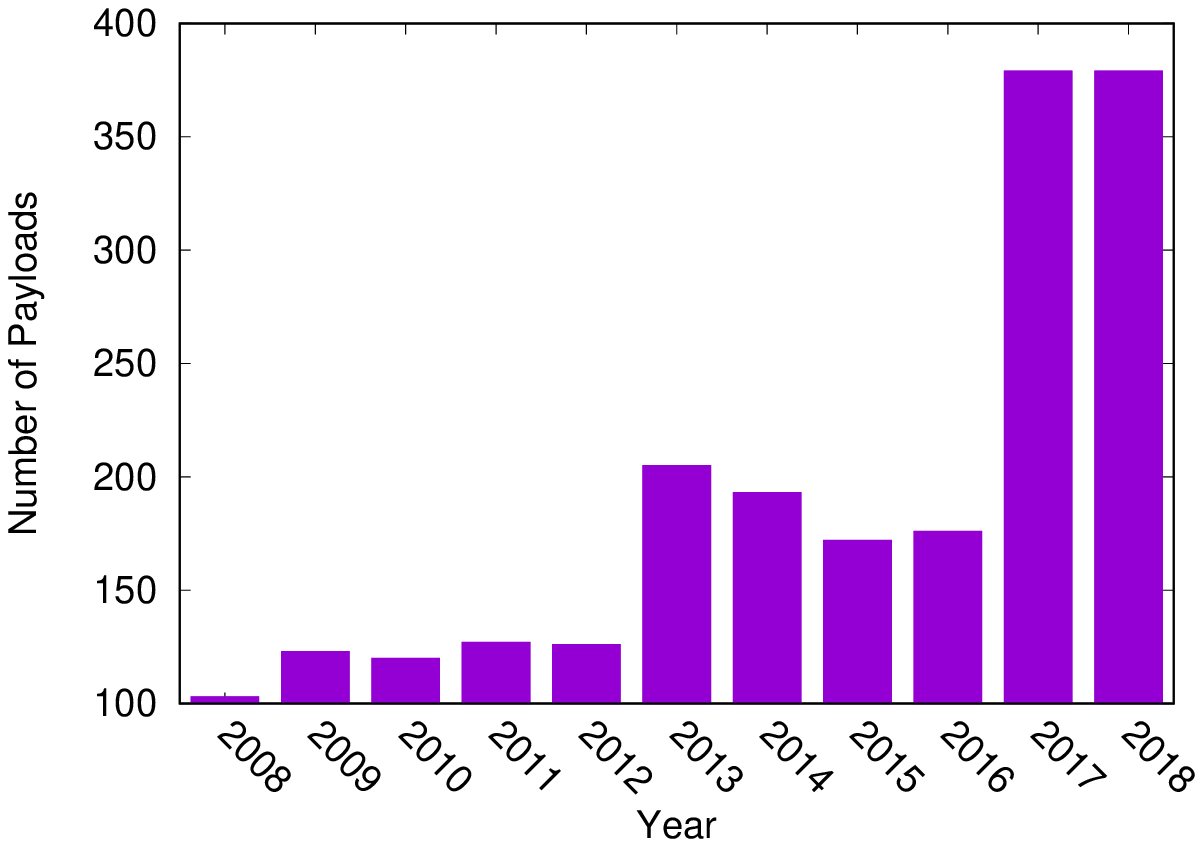}
       \caption{The number of payloads by date.  Data:  JSFCC.}
       \label{fig:payloads}
     \end{minipage}\hfill
    \begin{minipage}{0.45\textwidth}
      \includegraphics[width=3in]{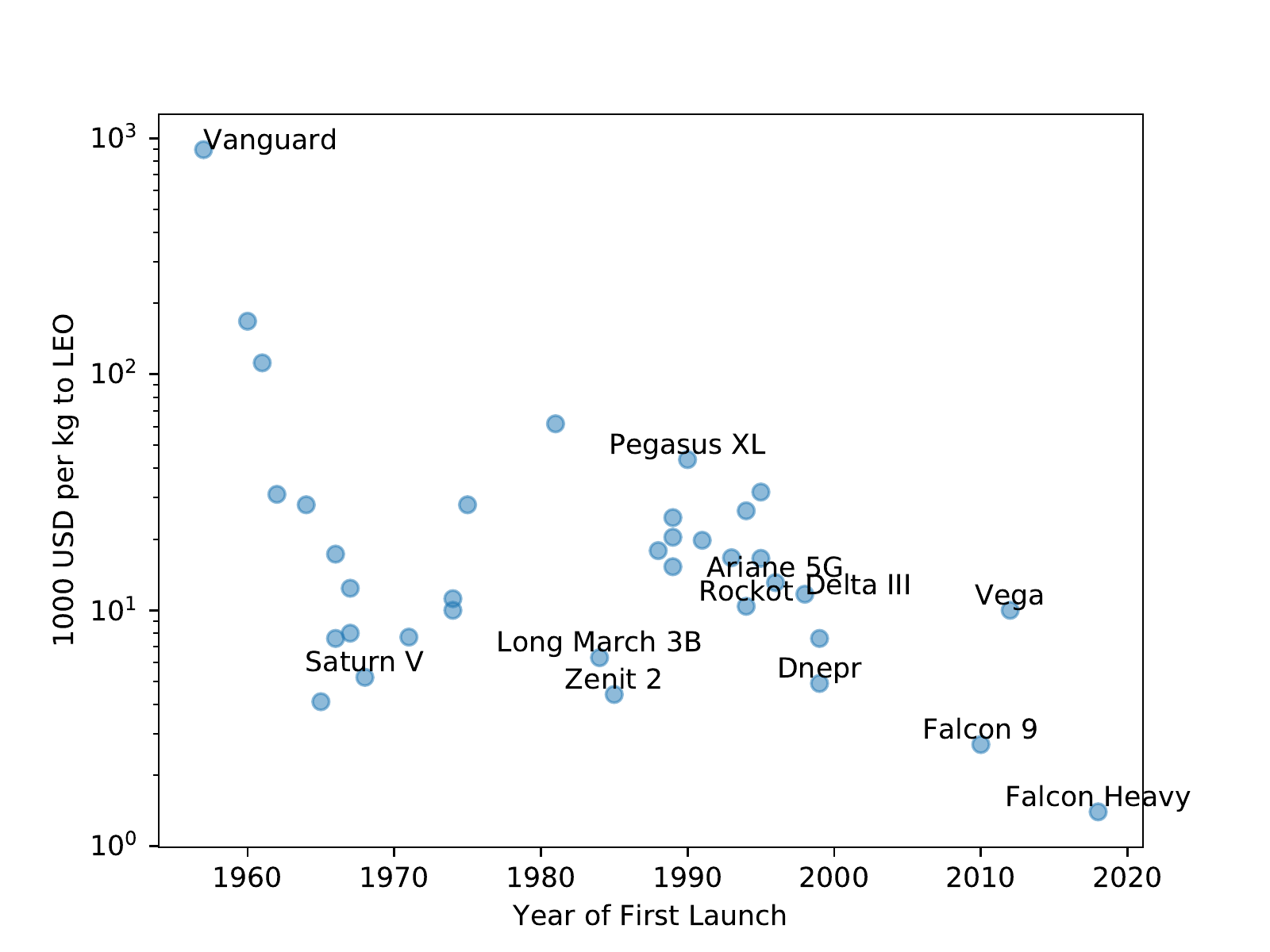}
      \caption{Cost to per kg to LEO, adjusted to 2018 dollars. Based on Jones~(2018). }
       \label{fig:launchcost}
    \end{minipage}
\end{figure}

Year 2018 established a new record for the number of orbital launches, with 114 in total\cite{spacelaunchreport}, 
up from 69 launches in 2008.  Over the past decade the number of payloads has also increased (Fig.~\ref{fig:payloads}). 
India alone carried 104 payloads in a single launch in 2017\cite{indiarecord}, almost all of which were cubesats. SpaceX launched 64 satellites at once in November of 2018.  Such ridesharing shows no signs of abating. In preparation for its megaconstellation Starlink, SpaceX placed 60 satellites, with a mass of 227 kg each, into orbit with one Falcon 9 in May of 2019.  As of August 1st this year,  209 satellites have already been launched, ensuring 2019 will be another high-payload year.

When comparing Figure \ref{fig:payloads} with the corresponding number of orbital launches for a specific year, most payloads are placed into orbit through ridesharing rather than dedicating a single rocket to a mission.   At least one company has emerged with the business model to be a liaison between end users and launch organizations\cite{spaceflight}, facilitating ridesharing. Several launch providers are  making ridesharing a priority for at least some fraction of their launches \cite{spacex}\cite{rocketlab}, while others are investing in new technologies to advance this service\cite{vegaSSMS}.  

For the number of orbital launches in 2018, China led with 39 from Chinese spaceports. In comparison, 31 rockets were launched by the United States, 17 by the Russian Federation, 11 by the EU, 7 by India, 6 by Japan, and 3 by New Zealand. 
Of the 31 launches in the US, 21 used  SpaceX\cite{spacex} rockets, 8 used  ULA\cite{ula}, and two by Orbital ATK.  Worldwide, there were 23 types of expendable rockets available for commercial use in 2017\cite{faa}.  

Falcon 9 launches are now reported to be under \$3,000 USD kg$^{-1}$, which represents a considerable decrease compared with the prices between 1970 and 2000 (see Figure \ref{fig:launchcost}). Note that launch costs below $\rm \$10,000~USD~kg^{-1}$ during that time are thought to be due to subsidies by the Chinese and Russian governments.   There are multiple reasons for the recent decline in costs (see Jones 2018\cite{jones2018} for a review). Reusable rockets are not yet the main driver of this reduction, with the near-term cost savings from rocket re-use still unclear and potentially only 30\%\cite{theverge}.  While launch costs are often evaluated in terms of price per mass of payload, several companies are now offering or plan to offer low-cost rockets for LEO and/or SSO with dedicated or rideshare capabilities, such as the Electron\cite{mann2017} (\$6M USD), the Terran\cite{terran} (\$10M USD), and the Firefly Alpha\cite{firefly} (\$15 M USD).  

With the decreasing costs of rockets and payloads to orbit, can Canada also develop lift capabilities?  Maritime Launch Services is a company, incorporated in Canada, formed from a partnership between Yuzhnoye, a Ukranian state-owned rocket manufacturer, and United Paradyne, an American private company with propulsion services.  An environment assessment was just approved, clearing a major hurdle for MLS to establish a commercial spaceport in Canso, Nova Scotia\cite{boucher2019}.
However, the upper stage of the proposed Cyclone-4M rocket would use unsymmetrical dimethylhydrazine (UDMH), which is known to have adverse health effects and has caused environmental damage in Russia and the polar regions\cite{byersbyers2017}.  
This raises concerns about the MLS's proposed solution. 

\begin{center}{\sc Spacecraft and the Orbital Debris Environment}\end{center}

{\it Space-based observatories will need to be cognizant of orbital debris, and all types of astronomical observations may become affected by megaconstellations.  }
\vspace*{0.5cm}

As of 7 August 2019, there are 19759 objects in orbit that are tracked and assigned international designations\cite{jsfcc,celestrak}.  This includes payloads (5143, of which approximately 2400 are active), rocket bodies (2196), and orbital debris (12315, excluding rocket bodies) with diameters $D\gtrsim10$ cm or larger.  The catalogue for this size range is thought to be incomplete\cite{esadebris}.  
Smaller debris particles are much more numerous, with an estimated 900,000 pieces with $D>1$ cm and over 100 million for $D>1$ mm\cite{esadebris}.  
At orbital speeds, millimetre-sized objects can cause local damage and disrupt spacecraft subsystems, while centimetre-sized objects can disable a satellite and cause explosions. 

\begin{figure}[h]
    \centering
    \begin{minipage}{0.45\textwidth}
    \includegraphics[width=3.5in]{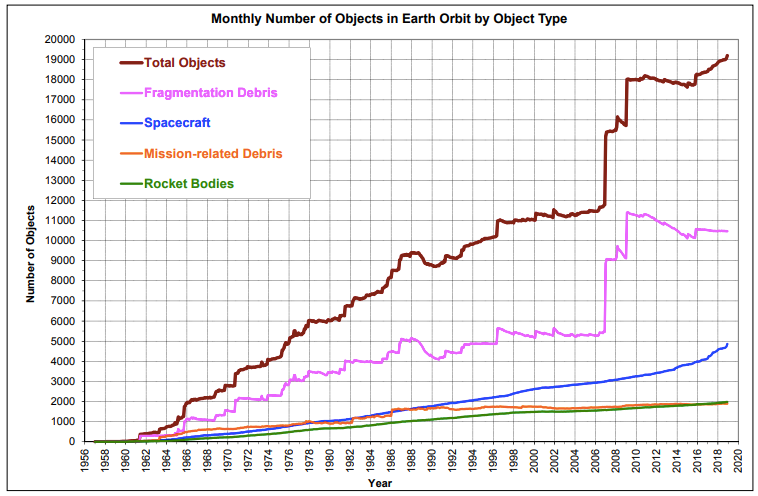}
    \caption{Debris and payloads in orbit as of May 2019. Fragmentation debris refers to breakup events, including intentional and unintentional explosions, while mission-related debris are objects that are not payloads or rocket bodies that are released while placing a payload into orbit. Image credit: NASA ODPO\cite{odqmay2019}. }
    \label{fig:orbitaldebris}
     \end{minipage}\hfill
    \begin{minipage}{0.45\textwidth}
      \includegraphics[width=3.25in]{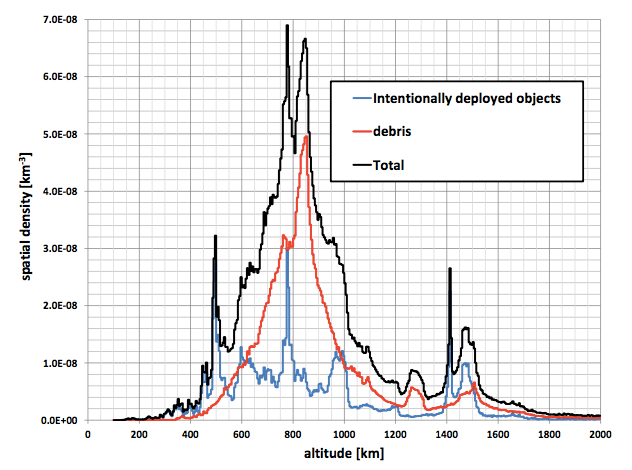}
      \caption{Spatial density of tracked objects in Earth orbits up to 2000 km. Image credit: Anz-Meador et al.~(2018)\cite{historicalbreakup2018}. }
       \label{fig:debrisdensity}
    \end{minipage}
\end{figure}

The accumulation of debris, rocket bodies, and payloads is shown in Figure \ref{fig:orbitaldebris}.  The 2007 spike in debris is due to the Chinese Anti-Satellite Test (ASAT), i.e., the intentional destruction of the Chinese-owned Fengyun 1C at an altitude of 890 km. At the time of this writing, there are 2818 catalogued debris particles still in orbit from this event.  The debris spike in 2009 is due to the Iridium 33 and Cosmos 2251 event, an accidental collision between an active and a defunct satellite, respectively.  Also occurring near an altitude of 770 km, there are still 1390 catalogued debris particles from this collision, increasing the risks to payloads operating at these altitudes (see Fig.~\ref{fig:debrisdensity}).  

As debris and new payloads accumulate, so grows the risk of the so-called Kessler Syndrome\cite{kessler1978}, which is a collisional runaway among Earth's artificial satellites and debris, similar to an astrophysical cascade\cite{dohnanyi1969}.  Operationally, the Kessler Syndrome is the state at which the main debris-generating mechanism is collisions between human-made objects in space.  While atmospheric drag will cause defunct satellites and debris to decay, this is only effective at low altitudes ($<2000$ km).
Nevertheless, decay timescales can still be long depending on the actual altitude and the area-to-mass ratio of the object. 
Future astronomy space missions will be operating in this environment.  
Megaconstellations may be particularly problematic for astronomy.

SpaceX was granted approval by the FCC in the United States to place approximately 12,000 satellites at different altitudes in LEO\cite{henry2019} to provide global internet. OneWeb, Telesat, and others are also proposing megaconstellations for telecommunications.
There will potentially be more satellites in the sky than naked-eye stars (the Yale Bright Star Catalogue lists 9110 stars). 
With more than one satellite on average every 3.4 sqr.~deg., megaconstellations  carry the potential to affect astronomical observations from radio through optical. 
Space-based mission profiles may need to be redesigned to be above communication shells, should those shells create unmanageable interference with the observations.  Even MEO is seeing a rise in orbit use due to states launching their own satellite-based navigation systems, such as GPS, GLONASS, BeiDou, Galileo, and NAVIC. 

The launch of megaconstellations is of further concern for the sustainable use of Earth orbital slots.  NASA's Orbital Debris Office has conducted several studies looking at hypothetical changes in the debris environment as the number of payloads in orbit increases by an order of magnitude\cite{odqaugust2019} over the next 20 yr.  The results depend on efficacy of post-mission disposal and a low satellite failure rate.  
Of the 60 Starlink prototypes launched in May 2019, three failed by June\cite{ocallaghan2019}. On September 2nd, 2019, Aeolus (operated by ESA) was manoeuvred to avoid a potential collision with Starlink 44 (SpaceX) during a predicted conjunction.   The situation was notable due to a communication failure between SpaceX and ESA\cite{esa_aeolus,cbc_aeolus}, which highlighted the lack of adequate space traffic management.  While still in its early stages, these failures highlight concerns for the management of megaconstellations and their impact on all space users.  

Finally, as we expand our activity to cis-lunar and lunar orbits, it must be recognized that there are few mechanisms that will lead to the natural de-orbiting of spacecraft and their debris in such an environment.  Removal through lunar impacts runs the risk of polluting the lunar surface and endangering lunar surface operations.

\begin{center} {\sc New Models for Spacecraft Operations} \end{center}

{\it Grant agencies and PIs will need to account for the roles of pay-for-use space-based astronomy providers in facilitating astronomical observations.  }
\vspace*{0.5cm}

The widely successful Canadian space telescope MOST\cite{most} was launched in 2003 and operated until its decommissioning in 2019 after a critical subsystems failure.   
The science conducted by the telescope is noteworthy, but so is its funding model during the spacecraft's last five years of operation.  In 2014, the CSA announced that it would discontinue funding\cite{mostCanadianPress}, and
shortly after, MSCI took over operations under a pay-for-use commercial model.  As of April 2018, MOST time could be purchased at a rate of \$6000 USD per week, a minimum amount to account for the  effort of tasking the satellite.  No time allocation committee (TAC) was required. While this created a barrier for general access to the telescope on a directly competitive basis, it allowed the telescope to continue to operate.   

Another example is a mission currently in development: Twinkle\cite{twinkle}. From its inception, Blue Skies Space Ltd.~has developed Twinkle under an astronomy user-service model, again guaranteeing access to space telescope time upon payment. No TAC will be involved. The advantage to this model is that the capital costs can be paid through a combination of government-industry investment grants, as well as private investors, while its operational costs will be collected from the users so long as there is demand for the facility. 

It is unclear whether science service business models will become common in the near future.  
However, with the continued growth of small and medium business participation in space, it is a scenario that should be taken seriously.  
NSERC grants are not optimized for this situation and could limit Canadian PI participation in such facilities.  

Science service models will also present an additional challenge: data proprietary periods.  Blue Skies Space Ltd., as of July 2018, is not developing a public archive and is intending to honour data ownership by the user in perpetuity.  While grant organizations could require the public release of data obtained through pay-for-use observations, a potentially significant amount of scientific value could be lost to observations taken outside of such requirements.  As an example, the number of archival HST papers, many of which are high impact, outpaces PI-driven papers\cite{whiteHST}.

\begin{center} {\sc Planetary Science, Space Exploration, and Space Astronomy in a Space Resource Economy} \end{center}

{\it A global effort among state and non-state actors is underway to access and use the Moon, as well as other celestial bodies. }
\vspace*{0.5cm}

On February 28th, 2019, Prime Minister Justin Trudeau announced that Canada is formally joining the United States in the development of the Lunar Orbital Platform-Gateway\cite{pmgateway}, with a proposed \$2 billion investment over the next 24 years.
It is unclear how the money will be allocated among related programs.
Gateway is an ambitious project with the aim to develop a space station in lunar orbit that would enable, for example, transfers between Earth and the Moon, lunar surface operations, and human habitation in lunar orbit for durations of a few months at a time.
The space station will require a high degree of autonomy, for which Canada is proposing to supply a sophisticated AI-driven robotic arm\cite{canadarm3}.
Along with this partnership is a new focus on commercial, scientific, and technological lunar activities.
While few details have yet to be released, the Federal Government has announced an initiative called the Lunar Exploration Acceleration Program\cite{leap}, with allocated funding of \$150M over 5 years. Earlier this year, the Future Lunar Exploration Activities call for ideas was released\cite{flea2019} (responses were due on 28 June 2019), which is interpreted as helping the CSA prepare for the formal launch of LEAP. 

Gateway is just one manifestation of a growing global desire among commercial actors to access the Moon and other celestial bodies. There are a number of reasons for this push, including scientific and commercial interests. 
Initiatives such as the Google Lunar XPRIZE sought to stimulate competition from private companies and organizations to land a payload on the Moon and to transmit images and video back to Earth.  The lunar XPRIZE ended in 2018 with no winners.  However, a private Israeli organization launched Beresheet, which crashed on the Moon's surface on 22 February 2019.  While the mission failed to safely land on the surface of the Moon, Beresheet became the first privately-funded spacecraft to touch the Moon's surface.  Google awarded the company the Moonshot prize as a result.

In 2014, the International Lunar Observatory Association (ILOA) selected Toronto-based Canadensys to place a small radio and optical observing payload in the south polar region of the Moon\cite{ilo1}.  Moon Express, an American company, was the other prime contractor. 
In 2018, the project announced a launch in late 2019, but the current status of the project is unknown.

In May 2019, NASA announced its selection of three small private companies to carry payloads to the Moon by 2020/2021 in preparation for a human surface operations in 2024\cite{moonsmallbusiness}.

While science and technology development are major drivers behind non-state actors' attempts to access the Moon, commercial interests are also a motivating factor, particularly for In Situ Resource Utilization.

\begin{center} {\sc In Situ Resource Utilization} \end{center}

{\it Future space missions will include the extraction and use of resources, such as metals, rare Earth elements, and water, from celestial bodies. }
\vspace*{0.5cm}

The Global Roadmap for Space Exploration\cite{roadmap}, developed by the International Space Exploration Coordination Group (ISECG), outlines mutual interests among 14 national space agencies, including the CSA.  
In the roadmap, in situ resource utilization (ISRU) is identified as a high priority among ISECG agencies due to its potential to lower exploration costs, extend mission durations, and open up mission opportunities that would otherwise not be possible. 
The idea is that space mining, whether on the Moon, asteroids, or other celestial bodies such as Mars, would provide the raw materials for local  spacecraft manufacturing, habitation, and other mission-related components.  Water is of particular interest, not just for human habitation, but for sourcing rocket fuel in space. 
Even the OSIRIS-REx mission, of which Canada is a 5\% partner, has mineral prospecting for ISRU as one of the motivations for the mission.

Commercial actors see the potential for a trillion dollar economy through ISRU.  
As a rough example, consider a C-type asteroid that is 100 metres in diameter, a density of $2000~\rm kg~m^{-3}$, and 10\% water by mass.  
Water could be extracted, in principle, by slowly breaking apart the asteroid and using solar radiation to release water and hydroxyl from minerals.  
Assuming this process is efficient, approximately $10^8~\rm kg$ of water could be harvested from the hypothetical asteroid.  
A SpaceX Falcon Heavy reusable rocket costs \$90M USD per launch, and is capable of delivering 26,700 kg to Geostationary Transfer Orbit\cite{spacex}, or about $\$3400\rm~USD$ per kg. 
Direct application of this cost to the hypothetical asteroid would suggest a worth over \$300B USD, although this does not accurately capture the economics.  The sudden availability of fuel in space would reduce the asteroid's value in non-trivial ways. The need to transport at least some of the resources further complicates the economic model.  Nonetheless, reducing the direct value by one or two orders of magnitude could still see ISRU as being commercially viable. 

Of interest for Canada is the recent publication (March 2019) released by Natural Resources Canada (NRCan) entitled ``The Candian Minerals and Metals Plan''\cite{NRCan}.  This document ``includes a vision, principles and strategic directions that governments, industry and stakeholders can pursue to drive competitiveness and long-term success''.  One of the Areas for Action under the section Science, Technology and Innovation is a recommendation that the federal government develop a policy approach for mining new frontiers (extreme climates, deep mining, offshore, and space) to foster investment and economic development.  It is indicated that specific Action Plans will be released based on the recommendations contained within the report, the first of which will be released in 2020. 

ISRU also has the potential to be of significant scientific value, as it could accelerate efforts to sample the composition of many lunar locations and asteroids.  It should nonetheless be recognized that the legal basis for space mining is controversial. Current law does not guarantee that commercial operators will preserve samples in a way that would be of general scientific value for planetary science; nor does it provide guidelines for providing samples and analyses of space resources to the scientific community.  

\begin{center} {\sc Changing Legal Landscape }\end{center}

{\it Nations are enacting domestic legislation that paves the way for ISRU.  }
\vspace*{0.5cm}

The 1967 Outer Space Treaty (OST) and four additional international treaties, including the Liability Convention, set the basis for Space Law.  An in-depth discussion of these treaties is out of scope for this WP; however, two areas of the OST are of particular importance.

Article I, paragraph 1 of the OST states that ``the exploration and use of outer space, including the Moon and other celestial bodies, shall be carried out for the benefit and in the interests of all countries, irrespective of their degree of economic or scientific development, and shall be the province of all mankind.''  Article II continues, ``Outer space, including the Moon and other celestial bodies is not subject to national appropriation by claim of sovereignty by means of use or occupation, or by any other means.''  Article VI, paragraph 1 connects the activities of non-governmental entities to the appropriate state (e.g., the state in which the company is incorporated). 

International space law does not clarify the extent to which, if at all, an entity can extract resources from a celestial body and claim ownership of those resources.  To provide legal certainty to investors, the United States Congress passed the U.S. Commercial Space Launch Competitiveness Act of 2015, which accords US citizens the right to own, possess, and sell space resources, at least under US law.  Luxembourg followed suit in 2017, while also offering financial incentives for companies to incorporate in Luxembourg.  Russia, UAE, and Japan are working on similar legislation.  Apart from NRCan noting that space is a new resource frontier\cite{NRCan}, Canada has yet to take a formal position

The sustainable development of space is a concern we now face.  How will state and non-state actors extract resources from celestial bodies, what oversight will be present, and what coordination will occur for the selection of sites or bodies?  What will be the environmental impacts to those bodies due to mining, and what are the unintended consequences for planetary science, planetary defence, and planetary astronomy?

%

\begin{center} {\sc Canadian-Led Deep Space Initiatives} \end{center}

{\it Leading deep space missions is within Canada's reach.}
\vspace*{0.5cm}

Canada's planned involvement with Gateway, as well as the increasing accessibility of space in general, presents an opportunity.  
Is it possible for Canada to lead deep space missions or even to be a major partner in the construction and operation of a lunar observatory?

Based on published values by SpaceX, an estimate of a translunar injection (TLI) payload is around 20t.  NASA's SLS Block 2 Cargo is expected to carry at least twice that to TLI, but the costs are unclear.  Using the Falcon Heavy as a base, the cost to send a payload to the Moon is around \$4500 USD per kg. Depending on the insertion into lunar orbit, costs could be somewhat higher.
If Falcon Heavy rockets are contracted to send supplies to Gateway, then this could enable small to moderate mass deep space missions to be placed into cis-lunar orbit. 
A 2t spacecraft could, in principle, be inserted into such an orbit for under \$10M USD. 
Such an opportunity would enable, for example, the placement of a relatively light-weight, 100 to 200 kg spacecraft with approximately $10~\rm km~s^{-1}$ of $\Delta v$ still available into cis-lunar orbit.  

Shipment of cargo to the lunar surface could also be enabled by Gateway.  Canada has significant investment and expertise in robotics and AI.
In addition to rovers, Canada could lead the placement of a lunar observing facility.  None of this is to suggest that Canada should do it alone; rather, Canada could be a majority partner rather than the typical 5\% or less as in most of Canada's past missions (e.g., see Hutchings 2019\cite{hutchings2019}).

Moreover, a Canadian-led deep space or lunar mission need not rely on Gateway.  In principle, a Falcon 9 configured to land the first stage on a drone ship could deliver just over 3000 kg to TLI\cite{launchperformance} at a cost of approximately \$43M USD (assuming a 30\% discount for the reusable configuration).  A mission profile similar to that used for Beresheet, which cost \$100M USD including launch\cite{beresheet_planetary}, could also be possible.  The mission used a GEO transfer orbit (GTO) with a Falcon 9, followed by orbital manoeuvres by the spacecraft\cite{spaceIL}. The Falcon 9 can deliver 8,300 kg to GTO. 

As an example mission concept, consider an on-demand asteroid rendezvous system (e.g., Beacon) built through the low-cost placement of multiple spacecraft into cis-lunar orbit. The spacecraft could be utilized to study near Earth asteroids during an apparition, a known low-$\Delta v$ opportunity, or a rapid response to a new discovery, a unique capability.  These rendezvous would enable the analysis of surfaces, sizes, and masses of multiple objects.  The spacecraft could also place beacons and/or retroreflectors on asteroids, providing long-term range monitoring for high-precision analyses of orbital changes due to the Yarkovsky effect, secular effects, and stochastic events such as surface changes or collisions with meteoroids.  On-demand rendezvous spacecraft would further make a first-of-its-kind contribution to planetary defence.  In particular, the need for countries to be willing and capable of providing asteroid rendezvous spacecraft during an impact emergency was brought into sharp relief during the Planetary Defence Conference 2019 tabletop exercise. Canada could lead such an effort. 

\newpage

\begin{center}Recommendations\end{center}

{\it Based on the material presented in the previous sections, we make the following recommendations.   }
\vspace*{0.5cm}

{\it Greater cooperation between the astronomical and the Space Situational Awareness communities is needed.} The increasing use of Earth orbits places stress on the debris environment and the safe operations of additional satellites.  Science spacecraft can both be threatened by and contribute to debris.  Partnerships between science and debris facilities could advance both communities.  NEOSsat was a step toward this aim and should be expanded.  Moreover, ground-based assets could further be utilized for astronomy and SSA, with the potential of significant cost sharing.  

{\it Build closer ties between the astronomical community and Global Affairs Canada (GAC).}  GAC establishes the Canadian regulations for utilizing space and for being the diplomatic conduit for negotiations relating to international space legislation and guidelines.  Based on current international activity, we expect substantial changes to space law over the next decade, including Canada taking a formal position on space mining and the use of the Moon. Involvement from the astronomical community now will help to shape Canadian legislation on space resource use and related activities, which could further have influence on international guidelines.   

{\it Establish a committee for evaluating the astro-environmental impacts of human space use, including on and around the Moon and other bodies. } In addition to promoting good stewardship of space, the astronomical and planetary science communities will be affected by orbital debris, megaconstellations, and the alteration of celestial bodies. While initiatives such as space mining could lead to major advances in sample returns or in situ analyses, there is no guarantee that private mining operations will preserve pristine astromaterials, will be interested in the same analyses as planetary astronomers, or will release what could be seen as proprietary information. Astro-environmental stewardship could also have direct planetary defence implications.  The proposed committee could further liaise with GAC and the SSA community. 

{\it CASCA and the Tri-Council should coordinate to identify programs that would enable Canadian astronomers to participate in pay-for-use services at appropriate funding levels.} MOST and Twinkle may be harbingers for a new approach to space missions.  Canadian astronomers should not be prevented from using space assets due to insufficient grant funds relative to their peers at foreign institutions. Because pay-for-use services may have extensive proprietary data models, Canadian astronomers should be required to make purchased data public on the Canadian Astronomy Data Centre after a short proprietary period.    

{\it CASCA should continue to foster a relationship with CSA, but also build close ties to the private space industry.} The variety of launch providers and options, the decreasing costs of launches, and the growth of commercial companies interested in space services may make it possible for the Canadian astronomical community to develop space missions without direct involvement of the CSA. 

{\it  Canadian-led deep space missions are within Canada's capabilities, and should be pursued. }  Gateway provides a path for partners to establish lunar operations, but also could make deep space accessible.  
Canada could take advantage of this situation to lead planetary science and deep space missions as a majority partner.  Even without Gateway, commercial launch capabilities enable ambitious mission planning, such as that carried out by the Beresheet mission.  A specific mission to pursue is designing, building, and launching asteroid rendezvous spacecraft tasked to study asteroid surface and bulk compositions and provide system for responding to planetary defence needs.  Such a mission could further place beacons on asteroids of high interest for long-term monitoring.



\begin{lrptextbox}[How does the proposed initiative result in fundamental or transformational advances in our understanding of the Universe?]

Many of the near-future transformational advances will require the use of space or will need to operate by observing through the growing space traffic.  The   recommendations here advocate for a proactive and multifaceted approach for addressing astronomy in the NewSpace era. 

\end{lrptextbox}

\newpage

\begin{lrptextbox}[What are the main scientific risks and how will they be mitigated?]

Multiple scientific risks are associated with NewSpace, most of which can be mitigated by collaboration with policy makers.  Inaction would lead to  space legislation that does not take into account the concerns and needs of the astronomy and planetary science communities.  Canadian science funding programs do not have a clear ability to enable Canadian PIs to use pay-for-service observing facilities, which may become prominent in the NewSpace era. Failure to engage in NewSpace further runs the risk of missed opportunities, such as leading space missions as a majority partner. 

\end{lrptextbox}

\begin{lrptextbox}[Is there the expectation of and capacity for Canadian scientific, technical or strategic leadership?] 

Canada has the opportunity to be a leader in advancing space policy that also promotes the sustainable development of space.  Canada can also take leadership positions in lunar activities and deep space missions, enabled by NewSpace infrastructure.  

\end{lrptextbox}

\begin{lrptextbox}[Is there support from, involvement from, and coordination within the relevant Canadian community and more broadly?] 

The Outer Space Institute uses a transdisciplinary approach to addressing issues in the sustainable development of space.  The institute was founded by a CRC in planetary astronomy, a CRC in global politics, and the former Chair of the UN Committee on the Peaceful Uses of Outer Space (COPUOS), who is Canadian.  The institute has direct involvement with the former president of the CSA, additional Canadian representation at the UN, and Canadian astronomers and planetary scientists. International involvement ranges from lawyers to social scientists to engineers to physical scientists.   

The McGill Institute for Air and Space Law addresses issues in space law, with a focus on legal perspectives.  They also have significant involvement with the Space Security Index, which is an international effort to identify security issues in space use using seventeen indicators. 

Canadian companies will be involved with Gateway, and Canadian industry is already involved in international efforts to operate on the lunar surface, independent of Gateway.  Additional examples are given in the WP.

Examples of international efforts to address NewSpace include COPUOS and its subcommittees, as well as the The Hague Space Resource Working Group.  

\end{lrptextbox}

\begin{lrptextbox}[Will this program position Canadian astronomy for future opportunities and returns in 2020-2030 or beyond 2030?] 

Failure to recognize the changing uses of space and to address new issues and opportunities arising from that use could hamper Canadian astronomy for decades. The recommendations discussed in this WP highlight future challenges that could affect all areas of Canadian astronomy.  

\end{lrptextbox}

\newpage

\begin{lrptextbox}[In what ways is the cost-benefit ratio, including existing investments and future operating costs, favourable?] 

Recommendations in this WP would create new committee positions tasked with building stronger connections with Global Affairs Canada and commercial space operators. Such a low-cost investment could have major benefits to the future of Canadian astronomy and planetary science.  

Canadian-led space missions, such as the Beacon mission proposed here, would require large investments, but could have a major impact on multiple areas of the Canadian space industry, with new opportunities for HQP.

\end{lrptextbox}

\begin{lrptextbox}[What are the main programmatic risks
and how will they be mitigated?] 

We refer the reader to the above responses and the WP, as there are multiple programmatic risks for individual sectors within NewSpace.  The main risks for  national and international governance issues are gaining receptive responses from policy makers.  This can be mitigated by establishing a strong advocate network.

Using Gateway infrastructure to advance Canadian space missions further carries risks based on partner dependencies.  This can be mitigated by exploring lunar and deep space mission options utilizing the Canadian space industry and adopting academic-commercial partnerships, which may include space resource utilization and space-based science service models.

\end{lrptextbox}

\begin{lrptextbox}[Does the proposed initiative offer specific tangible benefits to Canadians, including but not limited to interdisciplinary research, industry opportunities, HQP training,
EDI,
outreach or education?] 

The issues highlighted in this WP and the proposed initiatives are transdisciplinary, require government-industry-academic cooperation, and would expand opportunities for HQP.  

\end{lrptextbox}

\bibliographystyle{unsrt}
\raggedright
\footnotesize
\bibliography{example}

\end{document}